\documentclass[smallextended]{svjour3} %article} %[handout] %[colordvi]              
\title{What Are Observables in Hamiltonian Einstein-Maxwell Theory?\footnote{Forthcoming in  \emph{Foundations of Physics}}}  % 

\author{J. Brian Pitts} %\footnote}} 
\institute{ University of Cambridge,  jbp25@cam.ac.uk,  ORCID 0000-0002-7299-5137\\Funded by the National Science Foundation (USA) \#1734402 } % 

\date{\today}

\begin{document}

%%%%%%%%%%

\maketitle

\abstract{ Is change missing in Hamiltonian Einstein-Maxwell theory?  Given the most common definition of observables (having weakly vanishing Poisson bracket with each first-class constraint), observables are constants of the motion and nonlocal.  Unfortunately this definition also implies that the observables for massive electromagnetism with gauge freedom (Stueckelberg) are inequivalent to those of massive electromagnetism without gauge freedom (Proca).  The alternative Pons-Salisbury-Sundermeyer definition of observables, aiming for Hamiltonian-Lagrangian equivalence, uses the gauge generator $G$, a tuned sum of first-class constraints, rather than each first-class constraint separately, and implies equivalent observables for equivalent  massive electromagnetisms.  

For General Relativity, $G$ generates $4$-dimensional Lie derivatives for solutions. The Lie derivative compares different space-time points with the same coordinate value in different coordinate systems, like 1 a.m. summer time \emph{vs.} 1 a.m. standard time, so a vanishing Lie derivative implies constancy rather than covariance.  Requiring equivalent observables for equivalent formulations of massive gravity confirms that  $G$ must generate the $4$-dimensional Lie derivative (not $0$) for observables.  

These separate results indicate that observables are  invariant under internal gauge symmetries but covariant under external gauge symmetries, but can this  bifurcated definition work for mixed theories such as Einstein-Maxwell theory?  Pons, Salisbury and Shepley have studied $G$ for Einstein-Yang-Mills. For Einstein-Maxwell, both $F_{\mu\nu}$ and $g_{\mu\nu}$ are invariant under electromagnetic gauge transformations and covariant (changing by a Lie derivative) under $4$-dimensional coordinate transformations.  Using the bifurcated definition, these quantities count as observables, as one would expect on non-Hamiltonian grounds.    }


\section{Introduction}

%%%%%%%%%%%%%

When a Hamiltonian formulation of General Relativity was first undertaken  \cite{RosenfeldQG,SalisburySundermeyerRosenfeldQG}, the result   was expected  to be mathematically equivalent to the Lagrangian formulation.  Similar expectations held when Bergmann and his school reinvented comstrained Hamiltonian dynamics  \cite{AndersonBergmann}. Thus these authors employed a quantity, now called the gauge generator $G$, which combined the primary constraints (which express the impossibility of the Legendre transformation) and the secondary and later generations of the constraints (which are implied by the dynamical preservation of the primary constraints) in an essential way.  For example, in electromagnetism the gauge generator $G$ is $$\int d^3x [\pi^0 \dot{\xi}(x,t) - \pi^i,_i \xi(x,t)],$$ from which one quickly infers that $\{ A_{\mu}(t,y), G \} =  \xi,_{\mu},$ a familiar result.  (For the mathematical background, see \cite{Sundermeyer}.)

But soon novel Hamiltonian postulates were introduced that violated Hamiltonian-Lagrangian equivalence to facilitate merging GR with quantum mechanics.  Bergmann and Schiller postulated that the constraints act separately, not merely as a team, in producing gauge transformations   \cite[section 4]{BergmannSchiller}.  It was not long before the problem of time appeared:  observables were said to be  constants of motion \cite{BergmannGoldberg}.  In reaching this conclusion,  Bergmann and collaborators evidently assumed a similarity between electromagnetism (with its internal gauge symmetry) and GR (with its external gauge, that is coordinate, symmetry)   \cite{BergmannObservableNC,Bergmann} regarding  a $0$ Poisson bracket of observables under gauge transformations.   (It is sufficient for the vanishing Poisson brackets to be achieved using the constraints themselves, a condition known as  ``weakly vanishing.'')    Analogously,  Dirac  was so impressed by his  important  trivialization of the primary constraints  that he proposed shrinking the phase space  from $20 \infty^3$ to $12 \infty^3$ dimensions \cite{DiracHamGR}, dropping the momenta vanishing in the primary constraints and, more worrisomely, their canonical coordinates, winding up with the spatial rather than spatio-temporal metric.  That shrinkage obscured foliation-changing coordinate transformations and prevented consideration of the gauge generator $G$, which makes essential use of the primary constraints.

A key issue is whether  first-class constraints \emph{only as a team} generate changes of coordinates or other conventional redescription (gauge transformations)  \cite{RosenfeldQG,AndersonBergmann,CastellaniGaugeGenerator,PonsSalisburyShepleyYang}, or, as became the more popular view, does each first-class constraint \emph{by itself} generate a gauge transformation?  Both Bergmann \& Schiller's novel Hamiltonian postulates  and Dirac's shrinking the phase space pushed toward the separate first-class constraint view.  The separate first-class constraint view is supposed to be equivalent to the Lagrangian for ``observables'' \cite{HenneauxTeitelboim}.   But such equivalence depends upon a suitable definition of observables.  If this definition is itself postulated rather than derived, then the physical equivalence is itself merely a postulate rather than a result.   Such a postulate is not guaranteed to be consistent with more basic formulas.

Starting around 1980, the idea of recovering Hamiltonian-Lagrangian mathematical equivalence was revived, leading to the $3+1$ gauge generator $G$ that generates $4$-dimensional Lie derivatives  of the metric and its concomitants for solutions of Hamilton's equations  \cite{MukundaGaugeGenerator,CastellaniGaugeGenerator,SuganoGaugeGenerator,GraciaPons,SuganoExtended,PonsSalisburyShepley,ShepleyPonsSalisburyTurkish,PonsSalisbury,SundermeyerSymmetries,FirstClassNotGaugeEM}.   Barbour and Foster  also critique the claim that each first-class constraint generates a gauge transformation, albeit without embracing the gauge generator \cite{BarbourFosterPrimary}.

  Temporally overlapping with these reforms are some standard reviews (including some by Kucha\v{r}) describing the supposed absence of change in canonical quantum gravity \cite{KucharCanadian92,KucharCanonical93,IshamQuestionTime}.  Kucha\v{r}'s critique of the usual definitions helped to inspire the author's deviation from  the weakly vanishing Poisson bracket for observables in GR \cite{ObservablesEquivalentCQG,ObservablesLSEFoP}.  Gryb and Th\'{e}bault revise the definition of observables in a fashion more closely in line with Kucha\v{r}'s approach, but still quite distinct from it \cite{GrybThebaultSchrodinger}.   Anderson's extensive work also questions conventional definitions and massively extends work in the tradition of Kucha\v{r} and Barbour in many novel directions (\emph{e.g.}, \cite{AndersonProblemofTime}). 
An important question to consider is whether whatever problem of time actually exists, exists already at the classical level, or whether it appears at the quantum level after being resolved classically due to Hamiltonian-Lagrangian equivalence.

An advantage of the approach adopted here is that as far as possible, it avoids postulates and definitions about observables in favor of derivation from the Archimedean point of requiring equivalent observables for equivalent theories.  A limitation of the work thus far is its primarily classical character.

This paper will further explore a recent redefinition of observables, a redefinition built upon the gauge generator $G$ and the requirement that equivalent theories have equivalent observables---\emph{i.e.}, fixing or un-fixing the gauge (using the Stueckelberg trick or the like) does not alter the observables  \cite{ObservablesEquivalentCQG,ObservablesLSEFoP}.  This apparently novel principle (in the context of constrained Hamiltonian dynamics) vindicates the gauge generator $G$ over separate first-class constraints, but also requires a largely novel distinction between internal and external gauge symmetries (or something in that vicinity---see below), with invariance in the former case (including electromagnetism) and covariance (a tensor transformation law or the like) in the latter case (including gravity).  Thus  observables change by a $4$-dimensional Lie derivative, not $0$, under coordinate transformations, which are generated by $G$ for solutions of Hamilton's equations.  Requiring merely covariance, not invariance, under external (coordinate) transformation laws matches a conclusion drawn previously by consideration of the classical origins and meaning of the Lie derivative, especially the transport term \cite{GRChangeNoKilling}.  But this bifurcation raises the question whether mixed theories such as Einstein-Maxwell receive a consistent definition of observables. Will the $0$ and non-zero Poisson brackets conflict? The purpose of this paper is to show that the mixed definition (invariance for internal symmetries, covariance for external symmetries) indeed works for Einstein-Maxwell theory.

%%%%%%%%%%%%%%%%%%%%%%%

\section{Definitions of Observables}

Traditional conclusions involving the lack of change and being spatially global \cite{TorreObservable} have drawn criticism even  from general relativists without ties to the reforming Hamiltonian-Lagrangian equivalence literature.  Kucha\v{r} explicitly denies that observables should have $0$ Poisson bracket with what he takes to generate temporal gauge transformations, the Hamiltonian constraint $\mathcal{H}_0$ \cite{KucharCanadian92,KucharCanonical93}, though somehow he retains that condition for space with the momentum constraint $\mathcal{H}_i.$  (E. Anderson explores systematization of Kucha\v{r}'s ideas not in terms of space \emph{vs.} time but rather linearity \emph{vs.} nonlinearity \cite{AndersonObservablesBeables}; unfortunately, as Anderson notes, this does not work for supergravity.) 
Smolin's requirement that entities called observables be in fact observable in the ordinary sense by observers within the universe \cite{SmolinPresent} appears to conflict with the  $0$ Poisson bracket condition at least implicitly.   The failure of observables to play their expected role has also led to circumvention with new concepts \cite{RovelliPartialObservables,RovelliGPSObservables,DittrichPartialConstrained,TamborninoObservables}.  There are interesting similarities between these replacement concepts and the reformed definition of observables. Rovelli's partial observables are measurable but not predictable, whereas his complete observables are predictable.  For many  purposes predictability-up-to-gauge might suffice; the notion of observables that yields equivalence under gauge fixing involves predictability up to coordinate choice \cite{ObservablesEquivalentCQG}, implementing invariance under internal transformations (to which one cannot point) and covariance under external transformations (to which one can point).  It is striking that Rovelli finds that with test bodies, observables  include components of the metric tensor in a physically meaningful coordinate system, akin to Komar's conclusions \cite{KomarObservable} and not so different from the author's conclusion that the metric components (not referred to any special coordinate system:  covariant rather than invariant) are observable.

%%%%%%%%%%%%%%%%%%%%%%

\subsection{Bergmann \emph{vs.}  Bergmann on Observables}

It is not widely known that Bergmann was of several minds on observables. Indeed Bergmann seems not to have noticed the fact himself, but his ideas do not all fit together.   One widely recalled definition of his is that observables should have  (weakly) ${ 0}$ Poisson bracket with each separate first-class constraint  \cite{BergmannObservableNC,Bergmann}.  Given what follows from this definition, Kiefer rightly notes that such ``observables''  are a technical term, nonlocal, weakly tied to observation, and aimed at  quantum mechanics \cite{Kiefer3rd}.  On the other hand,  Bergmann  (sometimes) \emph{intended} otherwise, as one sees in little-attended works including his \emph{Handbuch der Physik} article: \begin{quote} General relativity was conceived as a local theory, with locally well defined physical characteristics.  We shall call such quantities \emph{observables}. \ldots We shall call \emph{observables} physical quantities that are free from the ephemeral aspects of choice of coordinate system and contain information relating exclusively to the physical situation itself.  Any observation that we can make by means of physical instruments results in the determination of observables;\ldots  \cite[p. 250]{BergmannHandbuch}.  \end{quote}
Such observables are  not constants of the motion and  do not require integration over the entire universe.  On occasion  Bergmann  wanted observables to be  independent of Hamiltonian formalism  \cite{Bergmann,BergmannHandbuch}\cite[p. 314]{BergmannKomarRoyaumont}, which would lead to Hamiltonian-Lagrangian equivalence.  Reading the bulk  of his work on observables \cite{BergmannObservableNC,BergmannHPA1956,BergmannChapelHill,BergmannLectures,BergmannLorentz,Bergmann,BergmannHandbuch,BergmannGeometryObservables,BergmannGoldberg,BergmannKomarRoyaumont,BergmannKomarStatus,BergmannKomarPRL,BergmannSchiller,BergmannNewman}, one suspects Bergmann was looking for a general relativistic analog of the transverse-traceless true degrees of freedom that one finds in electromagnetism \cite{ObservablesBergmann}.  Unfortunately nothing have most of those properties exists, though partial analogs, such as the use of transverse-traceless decompositions, are of course possible.  
  Given that Bergmann sometimes advocated views logically inconsistent with those often put forth on his authority, 
the tradition calls for discernment. The definition of observables considered below satisfies Bergmann's occasional preferences for Hamiltonian-Lagrangian equivalence and for spatio-temporally varying  observables.

%%%%%%%%%%%%%%%%%%%%%%%%%%%%%%%%%%%%%%%%%
 
 \subsection{Observables Reformed with the Gauge Generator $G$ } 

If gauge transformations are generated not by each first-class constraint by itself, but by a team $G$ of constraints working together by having interrelated coefficients (such as $8$ first-class constraints at each point but only $4$ arbitrary functions in vacuum GR \cite{CastellaniGaugeGenerator}, or in the electromagnetic case recalled above, $2$ constraints but only $1$ arbitrary function), then presumably the definition of observables should be reformed correspondingly, as Pons and collaborators have urged \cite{PonsDirac,PonsSalisburySundermeyerFolklore}.  
 Pons, Salisbury and Sundermeyer give an amended definition of observables by replacing  each first-class constraint with  gauge generator the $G$:   observables are gauge-{in}variant, having (weakly) ${ 0}$ Poisson bracket not with each first-class constraint, but with the gauge generator  $G[\xi^{\alpha}]$  \cite{PonsSalisburySundermeyerFolklore}. %$(\forall \xi^{\alpha})$

Recently the author showed that for massive electromagnetism, the requirement that equivalent theories have equivalent observables (in other words, that gauge-fixing/un-fixing doesn't change the observable content) is inconsistent with the separate first-class constraint view but fits perfectly with the gauge generator $G$ \cite{ObservablesEquivalentCQG,ObservablesLSEFoP}.  Massive electromagnetism approaches massless (Maxwell) as $m \rightarrow 0,$ whether classically or in quantum field theory \cite{BelinfanteProca,GoldhaberNieto2009,UnderdeterminationPhoton}.  It is a commonplace in quantum field theory that the de Broglie-Proca formulation without gauge freedom is useful for showing unitarity, whereas the Stueckelberg-Utiyama formulation with gauge freedom is useful for showing renormalizability   \cite[pp. 738, 739]{PeskinSchroeder}\cite[chapter 21]{WeinbergQFT2}\cite[chapter 10]{Kaku}. (One can view the de Broglie-Proca formulation as gauge-fixing the Stueckelberg field to $0$.)  Clearly the observables, at least on any definition that is worthwhile, are the same either way. Whatever the relationship between the empirical content of massive QED and the observables of classical Hamiltonian massive electromagnetism might be, it is equally clear that the non-gauge and gauge formulations must be equivalent. % The diagram illustrates the reasoning:

%\vspace{.13in} 
%
%%%%%%%%%%%%%%%%%%%%%%%%%%%%%%%%%%%%%%%%
%
% 
%{\bf Testing Definitions of Observables with Massive Electromagnetisms} 
%
%$$\begin{CD}
%\text{de Broglie-Proca}  @.    \text{Stueckelberg-Utiyama} \\
% \mathcal{L}: A_{\mu} \text{ observable}    @>install>gauge>  \mathcal{L}: A_{\mu}+\partial_{\mu} \phi  \text{ observable} \\
%% @.   @.  \\ 
%@VconstrainedVLegendreV                                       @VconstrainedVLegendreV \\
%% @.      @.  \\
%\mathcal{H}:  A_{\mu}, \pi^i \text{ observable} @>demand>equivalence>  \mathcal{H}:  \cancel{ \{O,FC\}=0 }  \\ 
% \text{because no FC constraints}      @.                \text{or } \{O,G\}=0. 
%\end{CD}$$

%%%%%%%%%%%%%%%%%%%%%%%%%%%%%%%%%%%%%%%%%%%%%%%%%%%%%%%%%%%%%%%%%%%%%%%%%%%%%%%%%%%%%%%%%%%%%%%%%%%%%%%%%%%%%%%%%%%%%%%%%%%%

\section{Observables and Internal \emph{vs.} External Gauge Symmetries, More or Less}

While the principle that equivalent theories should have equivalent observables vindicates the gauge generator $G$ over separate first-class constraints, there could be another distinction required between different types of gauge symmetries.  It is evident epistemologically that observables must be invariant under internal gauge transformations---they are inostensible, \emph{i.e.}, it is  impossible to point at an electromagnetic gauge choice or change thereof---so observable content cannot depend on such a choice.  Matters differ, however, with space-time coordinates and their transformations, which are familiar in daily life in Daylight Savings Time and in the work of geographers.   We can and do point to coordinate values and coordinate transformations routinely---a ball drops in New York at the start of the New Year, clocks are set forward an hour in the spring and back an hour in the autumn, and there is a golf course named for and located on the Prime Meridian near Cambridge, England.  With these conventions being accessible by pointing (ostensible), it  suffices for observables to be translatable from one set of conventions to another, much as natural languages are.  The transformation rules of tensor calculus, which yield the Lie derivative formulas, provide the translation manual. Hence covariance (translatability using tensor calculus) seems adequate.  

Invariance, on the other hand, is too demanding.  Because $G$ generates $4$-dimensional Lie derivatives, requiring invariance would imply that, for all vector fields $\xi^{\mu},$ $\{ O, G[\xi] \} =0,$ that is, that the Lie (directional) derivative of observable $O$ vanish along every vector field $\xi^{\mu}.$ The problem of spatio-temporal constancy is not resolved by using $G$.  The problem is not difficult to diagnose in terms of the meaning and derivation of the Lie derivative.  Unlike electromagnetic or Yang-Mills gauge transformations, coordinate transformations contain a transport term that compares the value of the field itself at two different space-time points.  For the space-time metric one has 
 $$ \pounds_{\xi} g_{\mu\nu} =  \left({ \xi^{\alpha}   \frac{\partial g_{\mu\nu}}{\partial x^{\alpha}} } + g_{\mu\alpha} \frac{\partial \xi^{\alpha}}{\partial x^{\nu}}  + g_{\alpha\nu} \frac{\partial  \xi^{\alpha}}{ \partial x^{\mu}}   \right);$$ while the second and third terms are analogous to Maxwell or Yang-Mills gauge transformations, the first term, the transport term, is totally different.  It  arises as the infinitesimal analog of comparing fields at  1 am Greenwich Mean Time  and 1 am British Summer Time (an hour apart). 
One compares different space-time points with the same coordinate values in different coordinate systems 
    \cite{WeylAction,KleinGREnergy1918,Noether} \cite[p. 271]{Landau} \cite{BergmannLectures}. Goldberg explains why this physically curious comparison is mathematically convenient 
  \cite[footnote 9]{GoldbergConservation}:  
\begin{quote} The $\bar{\delta}$  transformation compares the field variables at world points with the same coordinate value rather than at the same world point. That is, $\bar{\delta}y_A= \bar{y}_A(x)-y_A(x)= \delta y_A-y_A,_{\mu} \xi^{\mu}.$  The advantage of the $\bar{\delta}$ transformation is that it commutes with ordinary differentiation.
\end{quote} 
But clearly reality, observability, and gauge invariance do not require sameness at different events, even if one gives them the same coordinate value in different coordinate systems (which one can always do).  The changelessness of observables has arisen as a conclusion because it has been fed in as a premise through the (weakly) $0$ Poisson bracket condition in cases where $G$ generates a Lie derivative \cite{GRChangeNoKilling}.  Thus the changelessness of observables is resolved by imposing a more suitable requirement on $\{O, G[\xi^{\mu} ] \},$ namely,   $$\{O, G \} = \pounds_{\xi} O   \neq 0.$$ 
One might see Kucha\v{r}'s and Smolin's critiques of the usual definition of observables and Bergmann's occasional insistence on spatio-temporally varying observables as also pointing away from the $0$ Poisson bracket condition.

This definition  $\{ O, G \} = \pounds_{\xi} O $ can be rederived using the requirement that equivalent theories have equivalent observables.  One uses massive gravity, in one version without gauge freedom, in another version with gauge freedom  \cite{ObservablesEquivalentCQG,ObservablesLSEFoP}, one shows that the two empirically equivalent formulations have the same observables using the definition. % The diagram illustrates the reasoning: 
\vspace{.13in}

%{\bf Testing Definitions of Observables with Massive Gravities} 
%
%$$\begin{CD}
%\text{FMS Massive Gravity}  @.    \text{Parametrized Massive Gravity} \\
% \mathcal{L}: g^{\alpha\beta} \text{ observable}    @>install>gauge>  \mathcal{L}: g^{\mu\nu} X^A,_{\mu} X^B,_{\nu}  \text{ observable} \\
%%@.   @.  \\ 
%@VconstrainedVLegendreV                                       @VconstrainedVLegendreV \\
%%@.      @.  \\
%\mathcal{H}:  g^{\alpha\beta},  \text{ momenta }  @.   \mathcal{H}: \cancel{ \{O,FC\}=0 } \\ 
% \text{observable because }     @>demand>equivalence>                \text{or }  \cancel{ \{O,G\}=0} \\
%\text{no FC constraints }    @.                                   \text{or } \{O, G \} \sim \pounds_{\xi} O.
%\end{CD}$$

It turns out that ostensible \emph{vs.} inostensible, not internal \emph{vs.} external, is the fundamental distinction.  One sometimes sees a formulation of General Relativity with a background metric tensor and a non-coordinate gauge freedom, as well as a non-gauge coordinate freedom \cite{Grishchuk,SliBimGRG}.  One can combine a gauge transformation and a coordinate transformation to produce a transformation that changes only the background metric tensor, not the matter fields or the effective metric. This transformation involves the Lie derivative of the background metric $\pounds_{\xi}\eta_{\mu\nu},$ so one might think that it counts as an external transformation.  But because the background metric does not appear essentially in the field equations, it is unobservable.  Thus changes of only the background metric  leave all observables alone---that is, observables must be \emph{invariant} under such transformations.    Having a transport term in $\pounds_{\xi}\eta_{\mu\nu}$ is thus not the decisive factor.\footnote{  I thank Oliver Pooley for suggesting the possibility of seemingly `external' transformations (involving derivatives of the fields) for which invariance is nonetheless appropriate.} Fortunately paradigm internal transformations (Maxwell and Yang-Mills) and paradigm external transformations (coordinate transformations in General Relativity) do fit with invariance and covariance, respectively. 

%%%%%%%%%%%%%%%%%%%%%%%%%%%%%%%%%%%%%%%%%%%%%%%%%%%%%%%%%%%%%%%%%%%%%%%%%%%%%%%%%%%%%%%%%%%%%%%%%%%%%%%%%%%%%%%%%%%%%%%%%%%%%%%%%

\section{Mixed Internal-External Symmetry:  Einstein-Maxwell?}

One can now appreciate the importance of the question of how  theories with both internal and external gauge symmetries, such as Einstein-Maxwell, can receive a consistent definition of observables.  If one requires invariance of observables under both internal and external gauge transformations   \cite{PonsDirac,PonsSalisburySundermeyerFolklore}, then, I find,  for Einstein-Maxwell the electromagnetic field $F_{\mu\nu}$ is not an observable, because it is invariant under the electromagnetic gauge transformation but only covariant (changing by a Lie derivative $\delta F_{\mu\nu}= \pounds_{\xi} F_{\mu\nu}$) under a coordinate transformation.  This result follows by inspection from results on the Einstein-Yang-Mills theory \cite{PonsSalisburyShepleyYang}. (Whether one makes an extra electromagnetic gauge transformation to exclude unwanted velocities and thus render the formalism projectable to phase space is a matter of indifference in this respect, because such a transformation affects $A_{\mu}$ but not $F_{\mu\nu}.$)  Given that $F_{\mu\nu}$ is an observable given the definition for pure electromagnetism, one might be disappointed that $F_{\mu\nu}$ is not an Einstein-Maxwell observable on the definition requiring invariance under all gauge transformations.

 But given the bifurcated definition that requires internal (or rather, inostensible) invariance but external (or rather, ostensible) covariance \cite{ObservablesEquivalentCQG,ObservablesLSEFoP}, changing $F_{\mu\nu}$ by its Lie derivative, not by $0$ (covariance rather than invariance), is exactly what is required to make  $F_{\mu\nu}$ an observable in Einstein-Maxwell theory.  By  similar reasoning the space-time metric $g_{\mu\nu},$ which is observable on the bifurcated definition in vacuum General Relativity, remains observable in Einstein-Maxwell theory.
In more detail, on the bifurcated definition one wants   $\delta F_{\mu\nu}=0$ and  $\delta g_{\mu\nu}=0$ for electromagnetic gauge invariance, and  $\delta g_{\mu\nu}= \pounds_{\xi} g_{\mu\nu}$ and $\delta F_{\mu\nu}= \pounds_{\xi} F_{\mu\nu}$ from general relativistic coordinate  covariance.
 Fortunately the gauge generators for Einstein-Yang-Mills and their actions are  already known \cite{PonsSalisburyShepleyYang} and these results do in fact obtain, as one sees by inspection.  One simplifies Yang-Mills to Maxwell by dropping the  internal Yang-Mills index to reach Einstein-Maxwell ($A_{\mu}^i \rightarrow  A_{\mu}$), making the Yang-Mills structure constants disappear, and one takes the kinetic metric $C_{ij}=\delta_{ij}$ to be the number $1.$ 
Thus the mixed definition performs exactly as one would hope.  The electromagnetic field strength is an observable in Einstein-Maxwell just as it is in Maxwell's theory.  The space-time metric tensor is an observable in Einstein-Maxwell just as it is in GR.  The bifurcated definition behaves just as one would wish, unlike some other definitions.  

%%%%%%%%%%%%%%%%%

\section{Future Work: Local Supersymmetry?}

With the bifurcated definition behaving properly under internal, external, and combined internal-external definitions at least in key examples, it seems plausible that the definition works in most or all physically interesting cases, at least for theories lacking local supersymmetry.  But supergravity \cite{vanNReports}  poses  a  challenge in that the transformation rules for bosons and fermions evidently combine internal and external aspects in a non-diagonal way:  $\delta B \sim \bar{\epsilon}F$, $\delta F \sim (\partial B)\epsilon$. While Hamiltonian treatments have long been available (\emph{e.g.}, \cite{PilatiTetrad}), the gauge generators $G$ might not be known.  They are known, however, in $2+1$-dimensional supergravity \cite{Supergravity3dGaugeGeneratorMcKeon}.   Graviton-massive supergravity (a less desolate subject now than when there were reportedly $5$ extant papers only 15 years ago \cite{MassiveSupergravityDeconstruction}) might possibly permit resolution of the definition of observables by calculation as massive electromagnetism and massive gravity have.

%%%%%%%%%%%%%%%%%%%%%%%%%%%%%%%%%%%%%%%%%%%%%%%%%%%%%%%%%%%%%%%%%%%%%%%%%%%%%%%%%%%%%%%%%%%%%%%%%%%%%%%%%%%%%%%%%%%%%%%%%%%%%%

%\bibliography{Pitts}  
%
%\bibliographystyle{spmpsci}
%%%
%%%
%\end{document}     %apalike %unsrt
%

\end{document}